\documentclass[aps,twocolumn,prl,showpacs,preprintnumbers,amsmath,amssymb]{revtex4-1}
\usepackage{graphicx}
\usepackage{dcolumn}
\usepackage{bm}
\usepackage{natbib}

\begin{document}


\title{Predicted mobility edges in one-dimensional incommensurate optical lattices: An exactly solvable model of Anderson localization}


\author{J. Biddle}
\author{S. Das Sarma}
\affiliation{Condensed Matter Theory Center, Department of
Physics, University of Maryland, College Park, Maryland 20742, USA}



\begin{abstract}
Localization properties of non-interacting quantum particles in one-dimensional incommensurate lattices are investigated with an exponential short-range hopping that is beyond the minimal nearest-neighbor tight-binding model.  Energy dependent mobility edges are analytically predicted in this model and verified with numerical calculations.  The results are then mapped to the continuum Schr\"odinger equation, and an approximate analytical expression for the localization phase diagram and the energy dependent mobility edges in the ground band is obtained.
\end{abstract}


\maketitle
Anderson localization, the localization of electronic Bloch waves due to interference in disordered potentials, is one of the fundamental quantum phenomena in nature and is the transport mechanism behind metal-insulator phase transitions in solids \cite{Anderson}.  Although this mechanism was first proposed over 50 years ago, direct observation of Anderson localization has been notoriously difficult due to the problems in reliably controlling disorder in solid-state systems.  But recent advances in the manipulation of ultra-cold atoms offer a completely new, well-controlled tool in directly observing such fundamental quantum phenomena.  A notable example is the recent work done by Billy et. al. who observed Anderson localization in a diffuse Bose-Einstein condensate in a 1 dimensional (1D) waveguide with a disordered laser speckle potential \cite{Billy08}.  Another recent example is the work done by Roati et. al. who observed 1D Aubry-Andr\'e localization (a phase transition closely related to Anderson localization\cite{Aubry}) of cold-atoms in an incommensurate quasi-periodic potential\cite{Roati08}.  These advances highlight the potential of ultra-cold atoms to experimentally probe fundamental quantum localization phenomena that previously could only be studied indirectly or through numerical calculations.  Cold atomic systems offer precise control of the background potential, and the non-interacting limit is easily achievable with a very dilute gas of either bosons or fermions.  This is the context (and the motivation) of the current work, where we introduce a new and theoretically exact 1D localization model with mobility edges that should be observable in cold atomic systems.

1D localization phenomena are traditionally studied with the nearest neighbor tight binding model:
\begin{equation}
Eu_n = t_1(u_{n-1}+u_{n+1}) + V_nu_n, 
\label{eq:tbm}
\end{equation}
where $t_1$ is the hopping term representing tunneling between nearest neighboring sites and $V_n$ is the onsite disordered potential \cite{Anderson,Billy08} or the incommensurate potential \cite{Aubry, Roati08}. 
The simplicity of (\ref{eq:tbm}) allows for exact theoretical statements in certain cases.  For example, the 1D disordered Anderson model allows for only localized eigenstates at all energies independent of how weak the disorder may be \cite{Anderson}.  The Aubry-Andr\'e (AA) model with the 1D incommensurate potential has either all eigenstates extended or localized depending on the strength of the potential \cite{Aubry}.  Thus quantum localization in these 1D tight-binding models is, in some sense, trivial because all states are either localized or extended with no energy dependent localization transition as happens, for example, in the 3D Anderson model \cite{*[{Mobility edges in the 1D nn tight-binding model have been reported through numerical studies for certain multi-chromatic incommensurate potentials. See, for example, }] [{. In this Letter, we only consider the bichromatic problem}] Soukoulis82}.  
However, there is growing interest in exploring deviations from the tight-binding assumption \cite{Biddle, Boers07, *Moura, *Malshev, *Xiong, *Rodriguez03, *DasSarma88, *Riklund86}. Ultra-cold atoms loaded into optical lattices with controllable depths provide an experimental tool to study transport beyond the tight binding regime, where mobility edges are likely.  In this Letter, we introduce an exact analytically solvable 1D localization model which has a energy dependent mobility edge.  We believe that our model should be realizable in ultra-cold 1D atomic systems, and we show that our theoretical findings extend to the general Schr\"odinger equation description well outside the tight binding regime where cold atom localization experiments \cite{Billy08,Roati08} are typically carried out.

To highlight the new physics that may be observed with ultra-cold atoms in shallow lattices, we study localization in incommensurate lattices with an implicit short range rather than nearest neighbor hopping model.  In particular, we study the tight binding model:
\begin{equation}
 Eu_n = \sum_{n'\ne n} te^{-p|n-n'|}u_{n'}+V\cos(2\pi\alpha n + \delta)u_n, \label{eq:model}
\end{equation} 
where $\alpha$ is an irrational number, and $p>0$.  This is a simple exponential hopping generalization of the AA model.  We prove that this model has energy dependent mobility edges (contrary to the AA model), and we verify our analytical results by numerically diagonalizing (\ref{eq:model}).  We then show that our analytical results from this model can be used to predict the energy dependent mobility edges in the more fundamental Schr\"odinger equation for non-interacting particles in shallow, incommensurate optical lattices:
\begin{eqnarray}
(-\frac{\hbar^2}{2m}\frac{d^2}{dx^2}+\frac{V_0}{2}\cos(2k_Lx)+ \nonumber \\ \frac{V_1}{2}\cos(2k_L\alpha x))\psi(x)=E\psi(x),
\label{eq:SE}
\end{eqnarray}
where $V_0$ is the strength of the primary lattice and $V_1$ is the strength of the secondary lattice ($<V_0$).  We verify these results by numerically solving (\ref{eq:SE}) and examining the eigenstates.  The Schr\"odinger equation described by (\ref{eq:SE}) is often called the two-color or the bichromatic potential problem \cite{Roati08}.

\begin{figure}
\includegraphics[width=.47\textwidth]{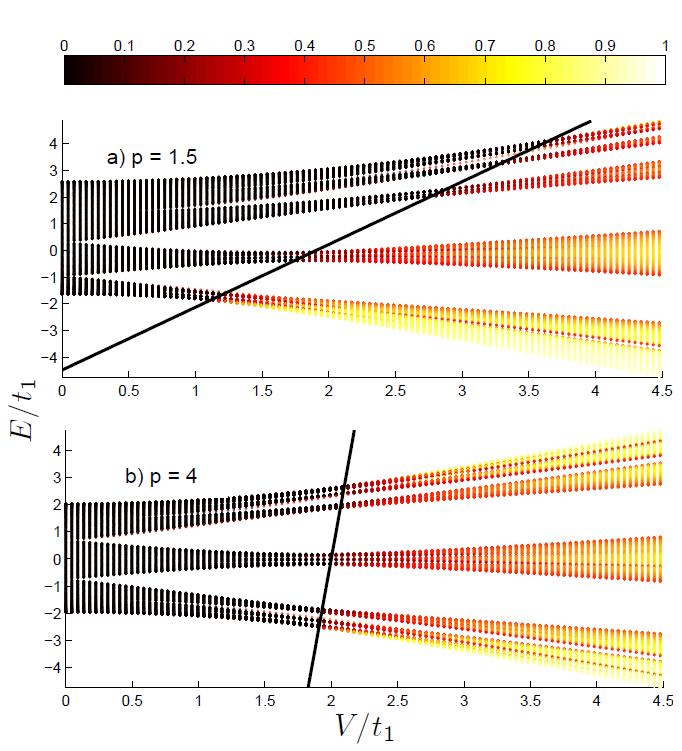}
\caption{\label{fig:specfigtbm} (Color online) Energy eigenvalues of (\ref{eq:model}) with 500 lattice sites and $\alpha = (\sqrt{5}-1)/2$ for a) $p=1.5$ and b) $p=4$.  The shading of the energy curves indicate the magnitude of the inverse participation ratio for the corresponding wavefunctions.  The solid line represents the analytical boundary between spatially localized and spatially extended states. }
\end{figure}
\begin{figure}
\includegraphics[width=.47\textwidth]{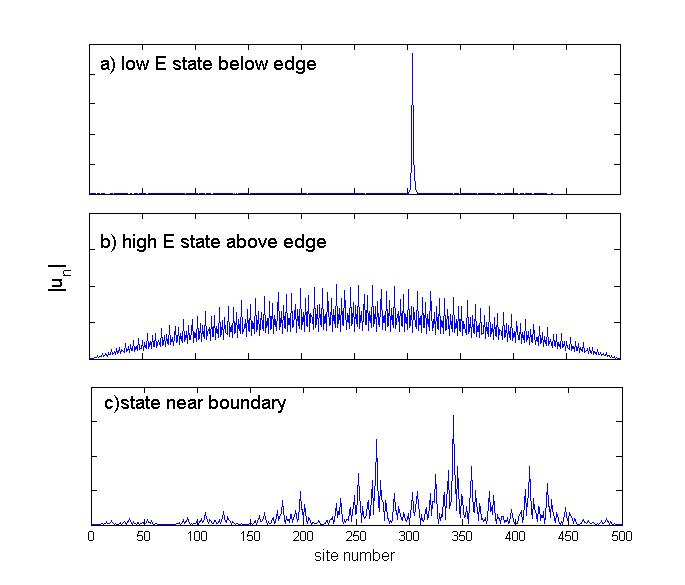}
\caption{\label{fig:statestbm} Eigenstates of (\ref{eq:model}) with 500 lattice sites, $\alpha = (\sqrt{5}-1)/2$, $V = 1.8$, and $p=1.5$ for different energy eigenvalues: a) low energy localized state below the mobility edge  b) high energy extended state above the mobility edge c) critical state near the mobility edge.}
\end{figure}
The AA model, where the potential in (\ref{eq:tbm}) is given by $V_n = V\cos(2\pi\alpha n + \delta)$, has been shown to be self-dual under the transformation:
\begin{equation}
u_n = \sum_{m} f_m e^{im(2\pi\alpha n+\delta)}e^{i\beta n},
\label{eq:AAtran}
\end{equation}
when $V = 2t_1$ \cite{Aubry}.  Since (\ref{eq:AAtran}) transforms spatially localized states into spatially extended states and vice-versa, it is argued in \cite{Aubry} that all eigenstates are extended for $V < 2t_1$ and localized for $V > 2t_1$.  The eigenspectrum at $V=2t_1$ has been shown to be singular continuous and the eigenvalues produce a Cantor-set structure \cite{Bellissard}.  This sharp transition between localized and extended states is often referred to as (AA) duality.  Obviously, the AA duality inherent in (\ref{eq:tbm}) does not apply to (\ref{eq:model}) and in particular, the transformation defined by (\ref{eq:AAtran}) does not work for the finite range hopping model.

We now show below that (\ref{eq:model}) allows for energy dependent duality points, (i.e a mobility edge).  Define the parameter $p_0>0$ such that:
\begin{eqnarray}
(E+t)-V\cos(2\pi\alpha n + \delta) = \omega^2T_n, \label{eq:param} \\
T_n = \frac{\cosh(p_0)-\cos(2\pi\alpha n + \delta)}{\sinh(p_0)}, \label{eq:Tn} \label{eq:w}
\end{eqnarray}
with $\omega^2 = \sqrt{(E+t)^2-V^2}$. It follows that $(E+t)/V = \cosh(p_0)$ and (\ref{eq:model}) can be cast in the form:
\begin{equation}
\omega^2T_nu_n = \sum_{n'}te^{-p|n-n'|}u_{n'}.
\label{eq:compact}
\end{equation} 
If we now consider the transformation:
\begin{equation}
\tilde{u}_m = \sum_{n}e^{im(2\pi\alpha n + \delta)}T_nu_n, 
\label{eq:tran}
\end{equation}
and note that for $p>0$ we have the identity,
\begin{equation}
T_n^{-1} = \sum_{m}e^{-p|m|}e^{im(2\pi\alpha n + \delta)},
\label{eq:Tnexpand}
\end{equation}
then it follows that the state, $\tilde{u}_m$ satisfies the equation:
\begin{equation}
\omega^2\tilde{T}_m\tilde{u}_m = \sum_{m'}te^{-p_0|m-m'|}\tilde{u}_{m'},
\label{eq:dualmodel}
\end{equation}
where $\tilde{T}_m$ is given by:
\begin{equation}
\tilde{T}_m = \frac{\cosh(p)-\cos(2\pi\alpha m + \delta)}{\sinh(p)}.
\end{equation}
We see that (\ref{eq:compact}) is self dual under the transformation (\ref{eq:tran}) when $p=p_0$.  Following AA \cite{Aubry}, we conjecture that all states are localized for $p>p_0$ and extended for $p<p_0$.  Then it follows that the condition for localization is given by the expression:
\begin{equation}
\cosh(p)=\frac{E+t}{V}.
\label{eq:condition}
\end{equation}
\begin{figure}
\includegraphics[width=.47\textwidth]{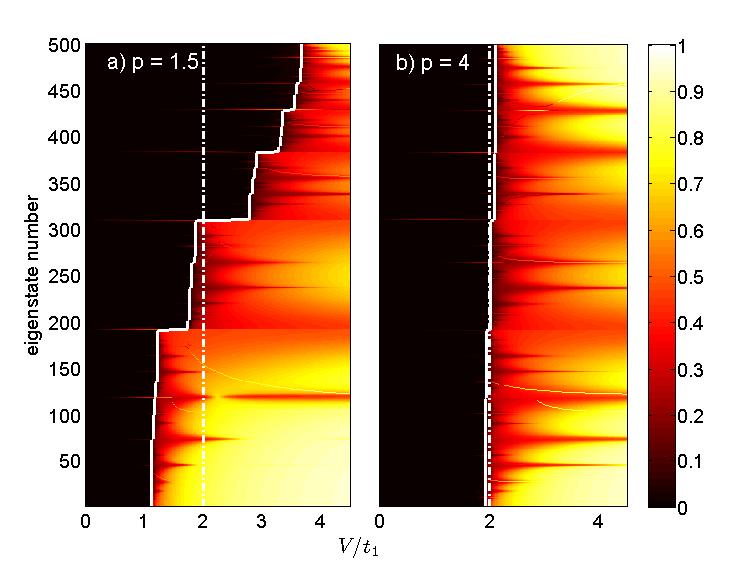}
\caption{\label{fig:IPRtbm} (Color Online) Inverse participation ratios of all eigenstates of (\ref{eq:model}) with 500 lattice sites and $\alpha = (\sqrt{5}-1)/2$ for a) $p=1.5$ and b) $p=4$.  The solid curves represent the analytical boundary between spatially localized and spatially extended states and the dashed lines represent the Aubry-Andr\'e condition.}
\end{figure}
It is straightforward to check that this condition becomes the AA condition where (\ref{eq:model}) becomes (\ref{eq:tbm}) in the limit $p\rightarrow\infty$, (i.e. $V=2t_1$).  The condition given by (\ref{eq:condition}) is the central new result of our work, showing that the model defined by (\ref{eq:model}) has an energy dependent mobility edge characterized by a transcendental equation.

To explicitly verify (\ref{eq:condition}), we numerically diagonalize (\ref{eq:model}) and study the spatial extent of the wavefunctions.  To do this we calculate the inverse participation ratio (IPR):
\begin{equation}
\text{IPR}^{(i)}=\frac{\sum_n|u^{(i)}_n|^4}{({\sum_n|u^{(i)}_n|^2})^2},
\label{eq:ipr}
\end{equation}
where the superscript $i$ denote the $i$-th eigenstate. The IPR of a wavefunction approaches zero for spatially extended wavefunctions and is finite for localized wavefunctions.

Fig.\ref{fig:specfigtbm} plots energy eigenvalues and the IPR of the corresponding wavefunctions for (\ref{eq:model}) as a function of potential strength, $V$, with $\alpha = (\sqrt{5}-1)/2$ and $p=1.5$ (Fig. \ref{fig:specfigtbm}a) and $p=4$ (Fig. \ref{fig:specfigtbm}b).  The solid line in the figure represent the boundary given in (\ref{eq:condition}).  As expected from our conjecture, IPR values are approximately zero for energies above the boundary and are finite for energies below the boundary, indicating that (\ref{eq:condition}) indeed defines the mobility edge for (\ref{eq:model}).   

Fig. \ref{fig:statestbm} displays eigenstates for $p=1.5$ and $V = 1.8$ at three different energy eigenvalues which lie above, below and near the mobility edge given by (\ref{eq:condition}).  We see that the wavefunction is localized for low energies (Fig \ref{fig:statestbm}a), extended for high energies (Fig \ref{fig:statestbm}b), and critical near the boundary (Fig \ref{fig:statestbm}c).
\begin{figure}
\includegraphics[width=.47\textwidth]{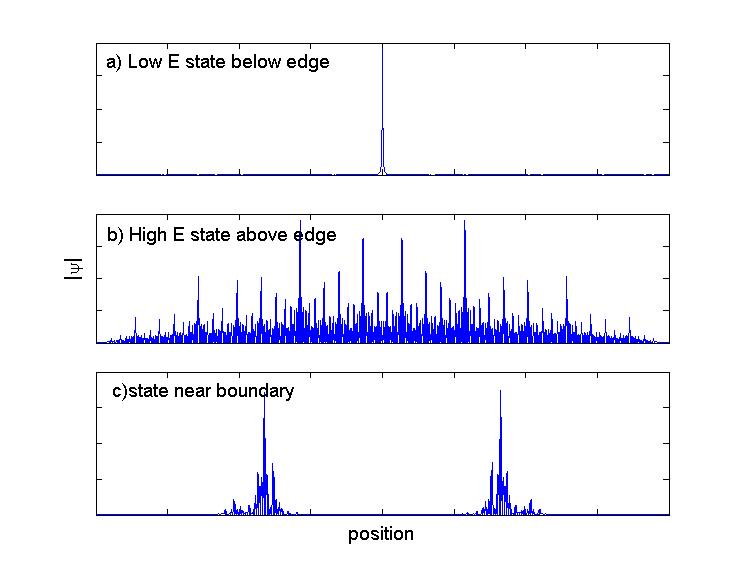}
\caption{\label{fig:statesse} Eigenstates of (\ref{eq:SE}) with 500 lattice sites, $\alpha = (\sqrt{5}-1)/2$, $V_0 = 2E_r$ and $V_1 =1.43E_r$ for different energy eigenvalues: a) low energy state below mobility edge  b) high energy state above mobility edge c) critical state near the predicted mobility edge.}
\end{figure}
In Fig. \ref{fig:IPRtbm}, we show the IPR of the wavefunctions as a function of eigenstate number, $i$, and potential strength, $V$.  The solid curves represents the predicted boundary given by (\ref{eq:condition}) and the dashed line is the AA self duality condition ($V=2t_1$).  We see in Fig. \ref{fig:IPRtbm} that our mobility edge prediction agrees well with the numerical IPR calculations.  In the case where $p$ is relatively small, the phase diagram shows clear mobility edges and differs markedly from the AA condition.  For large $p$, the slope of the localization condition is steep and is approximately equivalent to the AA condition, as expected.  

To understand the relevance of these results to ultra-cold atoms in optical lattices, we draw the connection between the exponential hopping tight-binding model given in (\ref{eq:model}) and the fundamental single particle Schr\"odinger equation in (\ref{eq:SE}).  To do this, we study the ground band Wannier functions \cite{Wannier}, $w_n(x)$ of (\ref{eq:SE}) for $V_1=0$ and approximate the matrix elements of the Hamiltonian in the Wannier basis.  Using the Gaussian approximation for the ground band Wannier states, the potential strength, $V$, in (\ref{eq:model}) is approximated by the expression:
\begin{equation}
V\approx \frac{V_1}{2}{\rm exp}(-\frac{\alpha^2}{\sqrt{V_0/E_r}}).
\label{eq:v}
\end{equation}
where $E_r\equiv(\hbar k_{L})^{2}/2m$ is the recoil energy.  Also from this approximation, we have for the constant energy difference between (\ref{eq:model}) and (\ref{eq:SE}):
\begin{equation}
E_0 = \langle w_n|H_0|w_n\rangle \approx \frac{1}{2}(V_0e^{-\sqrt{\frac{E_r}{V_0}}}+\sqrt{V_0E_r}), 
\label{eq:E_0}
\end{equation}
where $H_0$ is the Hamiltonian corresponding to (\ref{eq:SE}) with $V_1=0$.  The hopping coefficient, $t$, can be estimated using the deep lattice approximation for the ground bandwidth:
\begin{equation}
t\approx\frac{4}{\sqrt{\pi}}E_r(\frac{V_0}{E_r})^{3/4}{\rm
exp}(-2\sqrt{\frac{V_0}{E_r}}+p). \label{eq:t1}
\end{equation}
To estimate the exponential decay term, $p$, we make use of Kohn's results on the Kramers' function and its relation to the asymptotic behavior of the Wannier functions \cite{Kohn,Kramers}.  Using the deep lattice approximation for the effective mass at the top groundband edge of $H_0$, we obtain for $p$ the approximation\cite{Prodan, *[{See also, }] Fisch86}:
\begin{equation}
p\approx \cosh^{-1}(1+\frac{W_{1/2}}{2W_{0}}),
\label{eq:h0}
\end{equation}
where $W_0$ is the bandwidth of the ground band and $W_{1/2}$ is the width of the first bandgap.  The ratio $W_{1/2}/W_0$ can be estimated using properties of the Mathieu functions:
\begin{equation}
\frac{W_{1/2}}{W_0}\approx \frac {\sqrt{\pi}} {8} (\frac{V_0}{E_r})^{-1/4} \exp\left( 2 \sqrt{\frac{V_0}{E_r}}\right).
\label{eq:ratio}
\end{equation}
\begin{figure}
\includegraphics[width=.47\textwidth]{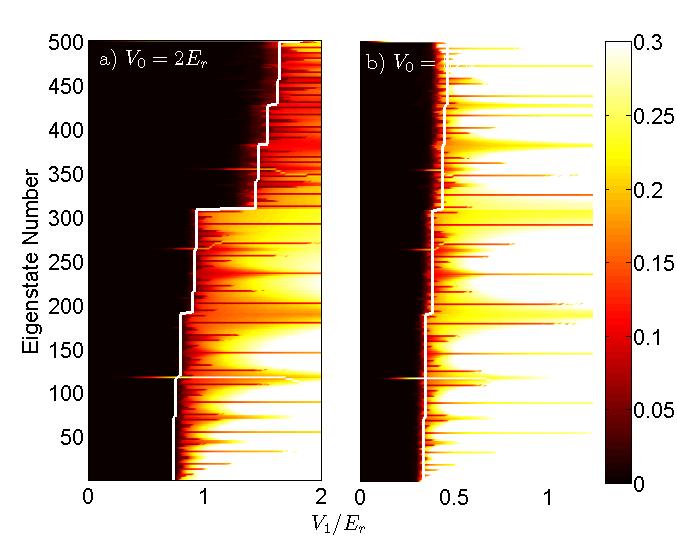}
\caption{\label{fig:IPRse} (Color Online) Inverse participation ratios of the approximate ground band eigenstates of (\ref{eq:SE}) with $\alpha = (\sqrt{5}-1)/2$ for a) $V_0=2E_r$ and b) $V_0=5E_r$.  The solid curves represent the analytical boundary between spatially localized and spatially extended states.}
\end{figure}
Finally, using (\ref{eq:h0}) and (\ref{eq:v}), the mobility edge for the 1-D incommensurate lattice Schr\"odinger equation, (\ref{eq:SE}), is given by:
\begin{equation}
2\exp\left(\frac{\alpha^2}{\sqrt{V_0/E_r}}\right)(E-E_0 + t) = V_1\left(1+\frac{W_{1/2}}{2W_0}\right),
\label{eq:SEcondition}
\end{equation}
where $E_0$ is estimated by (\ref{eq:E_0}).    
The condition in (\ref{eq:SEcondition}) is the Schr\"odinger equation equivalent of (\ref{eq:condition}).

To examine the accuracy of (\ref{eq:SEcondition}), we numerically integrate (\ref{eq:SE}) to obtain the energy eigenvalues and wavefunctions and calculate the IPR (obtained with (\ref{eq:ipr}) by replacing the sums with spatial integrals).  In our calculations, we set $k_L = 1$, $\alpha = (\sqrt{5}-1)/2$, $m = 1$.  The size of the system is given by $L = Na$ where $a$ is the lattice constant.  $N$ is chosen to be 500 and (\ref{eq:SE}) is sampled over 80,000 points.  Fig. \ref{fig:statesse} gives eigenstates at three different energy eigenvalues for $V_0 = 2E_r$ and $V_1 =1.43E_r$(similar to Fig \ref{fig:statestbm}).  Similar to the results in the tight binding model, we see that for a fixed potential strength an eigenstate can be localized for low energies (Fig \ref{fig:statesse}a), extended for high energies (Fig \ref{fig:statesse}b), and critical near the boundary (Fig \ref{fig:statesse}c).  Fig. \ref{fig:IPRse} gives calculated IPR values as a function of eigenstate number and $V_1$ for the first $N$ eigenstates (equivalent to the ground band when $V_1=0$)  The solid curves give the analytical boundary between localized and extended states as given by (\ref{eq:SEcondition}). We see in Fig \ref{fig:IPRse} that our analytical prediction is in good agreement with our IPR calculations. 
We also note that (\ref{eq:SEcondition}) is dependent on incommensuration and may predict no localization transition for $\alpha^2/\sqrt{V_0/E_r}\gg1 $, where the slope of the boundary in $E-V_1$ space is essentially flat.  This is consistent with numerical results reported earlier in \cite{Biddle} where localization transitions in (\ref{eq:SE}) are observed to be dependent on incommensuration for shallow lattices.
 
We have predicted the existence of an exact, analytic, energy dependent mobility edge in an incommensurate 1D model, which should be experimentally accessible in cold atomic systems.  This mobility edge is the energy dependent generalization of the AA self-duality concept and as such, the eigenstates precisely at the mobility edge are critical, (i.e. neither localized nor extended) with the mobility edge spectrum being singular continuous.  The fact that our predicted tight-binding mobility edge survives the continuum Schr\"odinger equation limit indicates that our exact prediction should be observable in cold atomic systems.  We note that we have explicitly numerically verified, \cite{Biddleunpub}, that the exponential hopping constraint is not crucial -- in fact, any bichromatic Schr\"odinger equation with hopping amplitudes falling off in some manner (e.g. power law, Gaussian, etc.) would approximately exhibit the predicted 1D mobility edges.  We also note that the slowly varying trap potential confining the cold atoms, not included in our calculations, should not affect our conclusions because it only acts as a finite size cut off for the extended states, which is incorporated in our numerical simulations.
\begin{acknowledgements}{ This work is supported by JQI-NSF-PFC and ARO-DARPA-OLE.  We thank Bin Wang and Donald Priour for helpful discussions.}
\end{acknowledgements}


%
\end{document}